\documentclass[aps,prb,twocolumn,superscriptaddress,showpacs]{revtex4}

\usepackage{amsmath}
\usepackage{graphicx}
\usepackage{subfigure}

\newcommand{\etal}{\textit{et al}. }

\begin{document}


\title{Possible High-Temperature Superconductivity in Hygrogenated Fluorine}


\author{D. A. Papaconstantopoulos}
\affiliation{Department of Computational and Data Sciences, George Mason University, Fairfax, Virginia 22030, USA}
\email[]{dpapacon@gmu.edu}

\begin{abstract}
Recent computational studies confirmed by experiment have established the
occurrence of superconducting temperatures, $T_c$, near 200 K when
the pressure is close to 200 GPa in the compound H$_3S$.
Motivated by these findings we investigate in this work the possibility of
discovering high-temperature superconductivity in the material H$_3$F.
We performed linearized augmented plane wave(LAPW) calculations followed by
the determination of the angular momentum components of the density of states,
the scattering phase shifts at the Fermi level and the electron-ion matrix
element known as the Hopfield parameter. Our calculated Hopfield parameters 
are much larger than those found in H$_3$S suggesting that they may lead to large
electron-phonon coupling constant and hence a large Tc similar or even larger
than that of H$_3$S. However, calculations of elastic constants are inconclusive
regarding the stability of this material.
\end{abstract}


\maketitle

\section{Introduction}

Back in the late sixties, Ashcroft\cite{PhysRevLett.21.1748} made
the bold prediction of room temperature superconductivity in
metallic hydrogen under very high pressures. Later in the seventies,
a quantitative evaluation of the electron-phonon (e-p)
coupling\cite{Ferro.16.307, PhysRevB.15.4221} using
the Gaspari-Gyorffy-McMillan (GGM) theories\cite{PhysRevLett.28.801,
  PhysRev.167.331} supported Ashcroft's ideas. In
Ref.\ \onlinecite{Ferro.16.307} an e-p coupling
$\lambda=1.86$ gave a superconducting transition temperature
$T_c=234$ K at an estimated pressure of 4.6 Mbar.

The ideas of Ashcroft have been recently confirmed by the
experiments of Drozdov \etal\cite{Drozdov2015} and a series of
theoretical
papers\cite{Duan2014,PhysRevB.91.184511,PhysRevB.91.060511,PhysRevLett.114.157004,flores,quan,li,bianconi}
that confirm hydrogen-based high-temperature superconductivity is
realized in the sulfur compound H$_3$S under 200 GPa
pressure. Reference \onlinecite{PhysRevB.91.184511} presents a
comprehensive set of calculations for H$_3$S using the GGM
theory. In a subsequent paper (Ref.15), we extended the work of Ref.8 studying substitutions of S by Si, P, and Cl
in the framework of the virtual crystal approximation. In the present paper we pursue another study in this
class of hydrides by substituting S by F.
So we have performed band structure and total energy calculations using the linearized augmented
plane wave(LAPW) method. The resulting angular-momentum components of the densities of states (DOS) at the Fermi level
($E_f$) and the phase shifts obtained from the computed band structure potentials are the input to the GGM theory for the evaluation of the Hopfield parameter ($\eta$).

\newpage

\section{Computational Details}

We have applied the LAPW code developed at NRL\cite{singh, nrl}, using the Hedin-Lunqvist form of
exchange and correlation, to calculate the band structure and total energy of the H$_3$F and H$_2$F systems
in the Im$\bar{3}$m and Fluorite crystal structures respectively.
The total energy minimization was done using the third-order Birch equation\cite{birch}. The total
and angular momentum decomposed densities of electronic states were
obtained by the tetrahedron method using LAPW results on a $k$-point uniformly distributed grid of 1785 k-points and
505 k-points for the respective irreducible Brillouin zones to ensure very accurate convergence. Subsequently, we
applied the Gaspari-Gyorffy (GG) formula to obtain the parameter
$\eta$, then the Allen-Dynes modification\cite{PhysRevB.12.905} of
the McMillan equation to determine $T_c$. The main steps here are to
determine the electron-phonon coupling constant $\lambda_j$ given by
McMillan\cite{PhysRev.167.331} as

\begin{equation}
\lambda_j = \frac{N(E_f) \langle I_j^2 \rangle}{M_j\langle\omega_j^2\rangle} \equiv \frac{\eta_j}{M_j\langle w_j^2 \rangle}
\end{equation}

\noindent where $N(E_f)$ is the total DOS per spin at $E_f$, $<I_j^2>$ is the electron-ion matrix element, $<w_j^2>$ is the
average phonon frequency and the index $j$ corresponds to hydrogen and fluorine.
 The Hopfield parameter $\eta_j$ for the two components is
computed by the GG formula shown below:

\begin{equation}
\eta_j = \frac{1}{N(E_f)} \sum\limits_{l=0}^2 2(l+1) \sin^2(\delta^j_l-\delta^j_{(l+1)})v^j_{l} v^j_{(l+1)}
\end{equation}

\noindent where $\delta^j_{l}$ is the scattering phase shift for the
$j$-th atom, the sum of which is related to the deformation
potential, and $v^j_{l}=N^j_{l}(E_f)/N^{j(1)}_{l}$ is the ratio of
the $l$-th partial DOS of the $j$-th atom to $N^{(1)}$, the free
scatterer DOS, for the given atomic potential in a homogeneous
system.The phase shifts $\delta^j_{l}$ are calculated using the following expression:

 \begin{equation}
 tan \	{\delta^j_l(R_s ,E)} = \frac{j_l^{'}(kR_s) - j_l(kR_s)L_l(R_s ,E)}{n_l^{'}(kR_s) - n_l(kR_s)L_l(R_s ,E)}
 \end{equation}
 
 where $L_l = \frac{u_l^{'}} {u_l}$ \  is the logarithmic derivative. 

\noindent The free scatterer DOS is defined as


\begin{equation}
N^{j(1)}_{l}=(2l+1)\int_{0}^{R_s} [u^j_{l}(r,E_f)]^2 r^2 dr
\end{equation}

\noindent where $u_l$ is the radial wave function and the upper
limit of the integral is the muffin-tin radius $R_s$. In previous
works, equations (2) and (3) contain multiplying factors of
$E_f/\pi^2$ and $\sqrt{E_f}/\pi$, respectively. But by examining
these equations it is easy to see that these factors cancel out.

Finally, we use the Allen-Dynes equation to determine the superconducting
transition temperature $T_c$ as follows:

\begin{equation}
T_c = f_1f_2 \frac{\omega_{\mathrm{log}}}{1.2} \exp{\bigg[-\frac{1.04(1+\lambda)}{\lambda-\mu^*(1+0.62\lambda)}\bigg]}
\end{equation}

\noindent In Eq.\ (4) we have set the Coulomb pseudopotential
$\mu^*=0.1$ and $f_2=1$. $f_1$ is the strong coupling factor given
by
\begin{equation}
f_1=\left[1+\left(\frac{\lambda}{2.46+9.35\mu^*}\right)^{1.5}\right]^{1/3} ~ .  
\end{equation}
It turns out for this material, $f_1$ can provide an additional
10\% enhancement to $T_c$. We have used the values for
$\omega_{\mathrm{log}}$ and $\langle \omega_j^2 \rangle$ found in
Ref. \onlinecite{PhysRevB.91.184511} from the analysis of the
results of Duan \etal (Ref.\ \onlinecite{Duan2014}). Our choice of
$\mu^*=0.1$ can be justified by the empirical formula proposed by
Bennemann and Garland\cite{bennemann}. 

\section{Results}

In Fig.\ 1 we show the Pressure v.\ Volume  relationships found from the Birch fit for the H$_3$S and H$_3$F compounds. It is worth noting that there is a significant difference between the two graphs showing that the H$_3$S reaches the pressure of 200 GPa  at much higher volume than in H$_3$F. So at $V=87.8$ (lattice constant $=5.6$ Bohr) the pressure is around 210 GPa in H$_3$S while at the same volume H$_3$F reaches a pressure of only 82 GPa. This suggests that H$_3$F might reach high superconducting temperature at much lower pressure than H$_3$S.

\begin{figure}[!htb]
\centering
\includegraphics[width=2.18in, angle=-90]{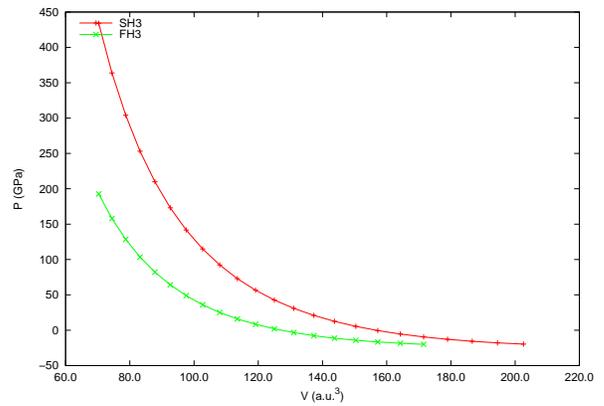}
\caption{ Pressure v.\ Volume relationships for H$_3$S and H$_3$F.}
\end{figure}

Fig.\ 2 displays the energy bands of H$_3$F in the bcc-like Im$\bar{3}$m structure for lattice constant $a=5.6$ Bohr  ($P=82$ GPa). We note that the low energy band near -1.0 Ry is almost 100 per cent of s-like fluorine character.
At the Fermi level, $E_f$, at about 0.9 Ry the bands consist of 70 per cent p-like fluorine character ,22 per cent
hydrogen s-like, 5 per cent fluorine s-like and 3 per cent fluorine d-like. Our Birch fit found that P=0 corresponds
to a lattice constant of 6.33 Bohr.

\begin{figure}[!htb]
\centering
\includegraphics[trim={0 0 0 0.4cm},clip,width=3.11in]{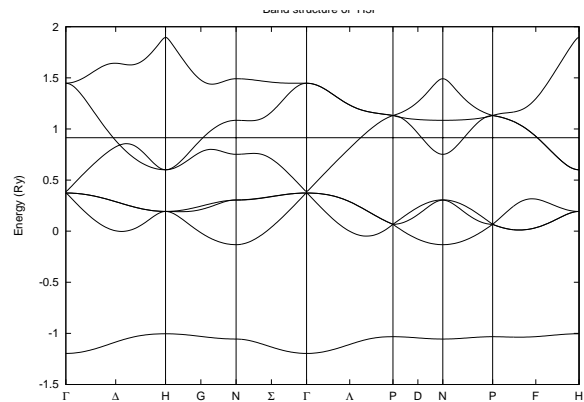}
\caption{Energy bands of H$_3$F for lattice constant  $a=5.6$ Bohr\ ($P=82$ GPa).}
\end{figure}
 
In Fig.\ 3 we present the total and angular momentum and site-decomposed(DOS) for H$_3$F in the Im$\bar{3}$m structure for lattice constant $a=5.6$ Bohr . We note the narrow s-like fluorine dominated peak at -1.0 Ry. This is followed by a gap of about 1 Ry where two fluorine dominated p-like peaks appear. Then at an energy of 0.5 Ry a tiny gap is found
which is followed by another two peaks with both fluorine p-like and hydrogen s-like contributions. In the middle of the latter two peaks $E_f$ is found. The $N(E_f)$ is decomposed as discussed above in the description of the bands. It is important to state here that the overall features of the DOS shown in Fig. 3 are very different from those calculated by many groups for H$_3$S. But at $E_f$ both the DOS values and the per site decomposition are very similar. 

\begin{figure}[!htb]
\centering
\includegraphics[width=2.18in,angle=-90]{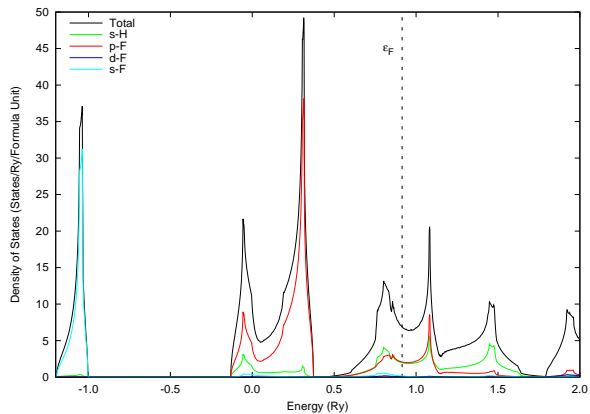}
\caption{Total and angular momentum-decomposed DOS for H$_3$F. Although this DOS has a different overall shape than that of H$_3$S, 
it turns out that at the Fermi level both the actual values and the decomposition are very similar between the two compounds}
\end{figure}
 
In Fig.\ 4 we show the values of the Hopfield parameter $\eta$ comparing H$_3$F to H$_3$S. The results shown in this 
figure establish a dramatic increase of the fluorine component of $\eta$ in H$_3$F over the corresponding value of the sulfur component in H$_3$S while the hydrogen component is comparable to that in H$_3$S.
More specifically from Fig.\ 4 we can see that at $P=128$ GPa (lattice constant $a=5.4$ Bohr) and for $P=82$ GPA (lattice constant $a=5.6$), the corresponding values of the $\eta$ fluorine are 17.5 eV/\AA$^2$ and 13.9 eV/\AA$^2$ respectively. As can be seen from the figure these values are almost a factor of three larger than those of both the sulfur and hydrogen components in H$_3$S which are actually achieved at higher pressures. This 
large increase of the parameter $\eta$ in H$_3$F is a signal that we should be looking for a high superconducting transition temperature in this compound if it can be synthesized.

\begin{figure}[!htb]
\centering
\includegraphics[width=3.11in]{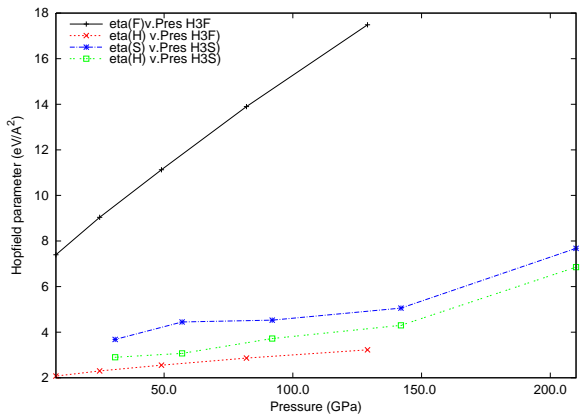}
\caption{Comparison of the Hopfield parameters $\eta$ as a
  function of pressure for H$_3$F and H$_3$S. Note that the values for the hydrogen components have been multiplied by three.}
\end{figure}
 
However, in order to obtain a quantitative prediction of the transition temperaturerge $T_c$, a large value of the 
Hopfield parameter is not a sufficient condition. It is necessary to estimate the force
constants $(M\omega^2)_j$ so that values for the electron-phonon
coupling constants $\lambda$ can be obtained. Using our previous
analysis\cite{PhysRevB.91.184511} for pure H$_3$S and the results of
Duan \etal\cite{Duan2014}, we derived the following values of the
averaged phonon frequencies in H$_3$S: $\langle\omega\rangle_S=615$K,
 $\langle\omega\rangle_H=1840$ K, and $\omega_{\mathrm{log}}=1560$K.
Now we assume that the $M\omega^2$ of H (optic mode) to be nearly the same as in H$_3$S.
We then estimate the $M\omega^2$ of the fluorine site by scaling the H$_3$S results by
the fluorine mass also introducing a volume dependence by considering the square of the
phonon frequency as proportional to the bulk modulus $B$. 
Hence, as shown in (Eq.1), by dividing our calculated parameters $\eta$ by the above
estimated values of  the force constants we obtain an estimate of $\lambda$ which is
shown as a function of pressure in Fig.\ 5.

\begin{figure}[!htb]
\centering
\includegraphics[width=2.18in]{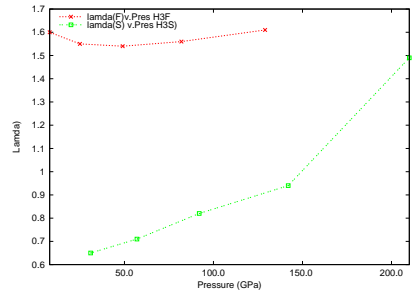}
\caption{Comparison of the electron-phonon coupling constants $\lambda$ as a function of pressure for H$_3$F and H$_3$S}
\end{figure}

Finally, using the Allen-Dynes equation (Eq.5) we calculated the superconducting transition temperature $T_c$. 
This estimate of $T_c$ for H$_3$F together with that of H$_3$S are shown in Fig.\ 6. It is interesting that for
the fluorine compound we predict transition temperature well over 200K for a pressure of only about 130 GPa.

\begin{figure}[!htb]
\centering
\includegraphics[width=2.18in]{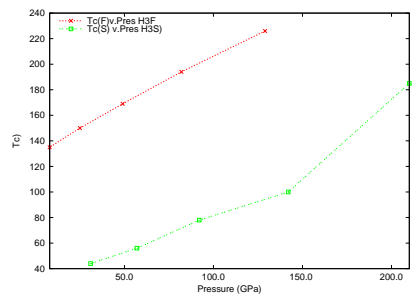}
\caption{Comparison of the superconducting transition temperature $T_c$ as a function of
 pressure for H$_3$F and H$_3$S}
\end{figure}

\section{Further Discussion}
We now proceed with further analysis of our results. The main result of our calculation
is the finding that the fluorine component of the Hopfield parameter $\eta$ is very
large in H$_3$F (see Fig.\ 4).
This is due to the very large contribution from the pd channel of F in the GG formula (Eq.3),which has the value of 13.7 eV/\AA$^2$ and 11.3 eV/\AA$^2$ for a=5.4 a.u. and a=5.6a.u.respectively. It is worth noting in H$_3$F the hydrogen component of $\eta$ is much smaller than in H$_3$S. 
In summarizing the situation we recognize that while our $\eta$ calculations are reliable, our estimates of the force constants are less reliable since we have not calculated 
the phonon frequencies from first principles. Nevertheless, the large values of $\eta$
are very intriguing especially since they are not due to large value of N(Ef) which has modest values of less than 7 states/Ry. Further support for the large $\eta$ is found 
from a calculation we performed in the Fluorite structure compound H$_2$F where we find
even larger values of $\eta$ exceeding 27 eV/\AA$^2$.
Therefore, it becomes important to check the stability of H$_3$F by calculating the
elastic constants c11-c12 and c44. We performed such calculations for the lattice constants a=5.4 a.u. and a=5.6 a.u which correspond to the highest pressures we considered.
The results are shown in Fig.\ 7 which depicts the energy versus the square of the distortion for c44 and c11-12.

\begin{figure}[!htb]
\centering
\subfigure[]{
\includegraphics[width=2.18in]{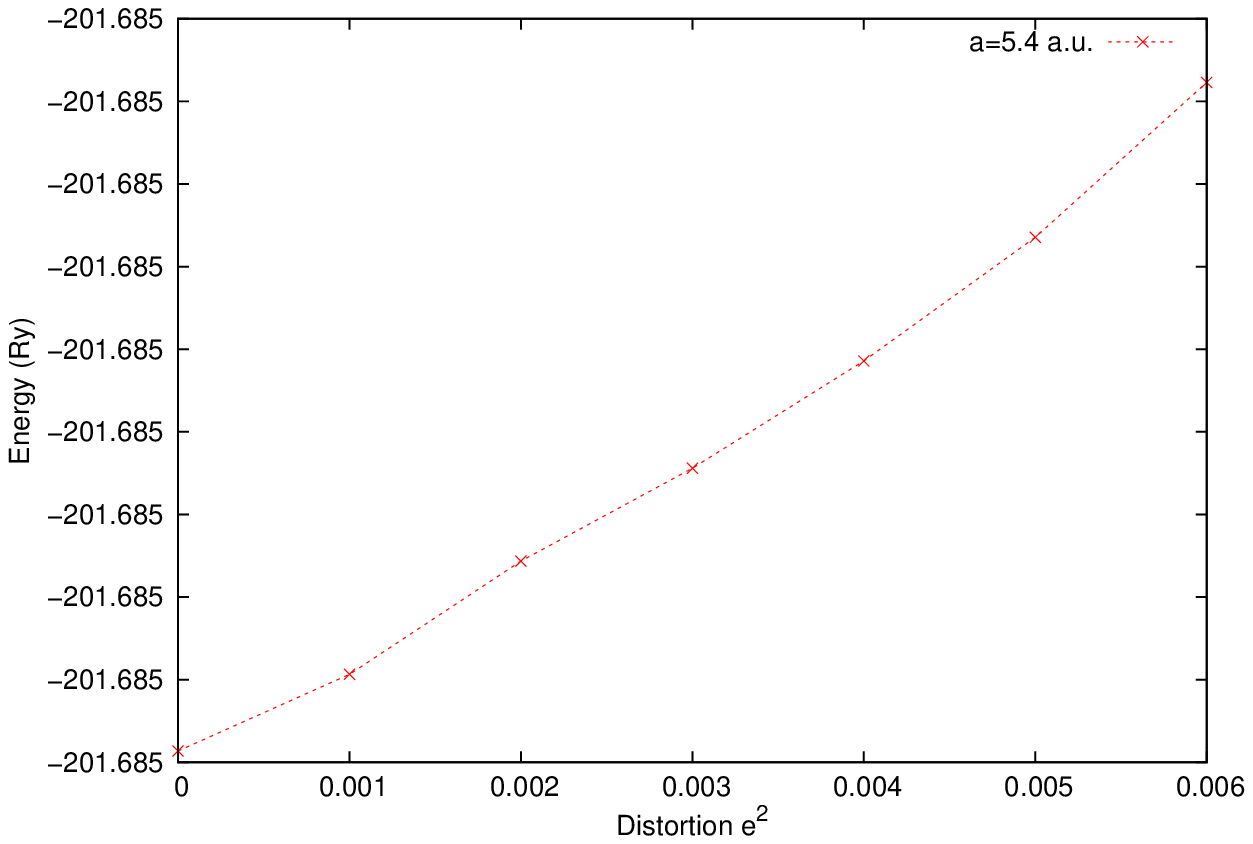}
}
\subfigure[]{
\includegraphics[width=2.18in]{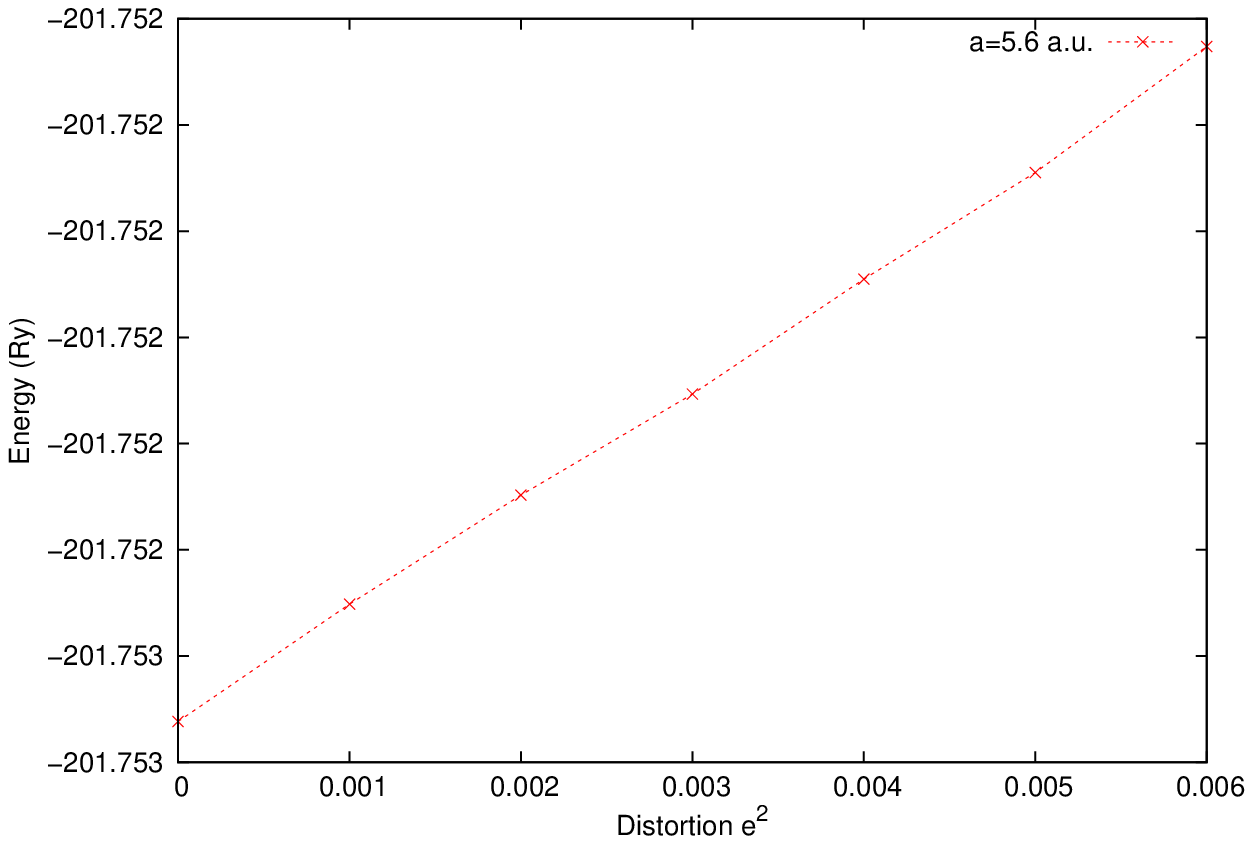}
}
\subfigure[]{
\includegraphics[width=2.18in]{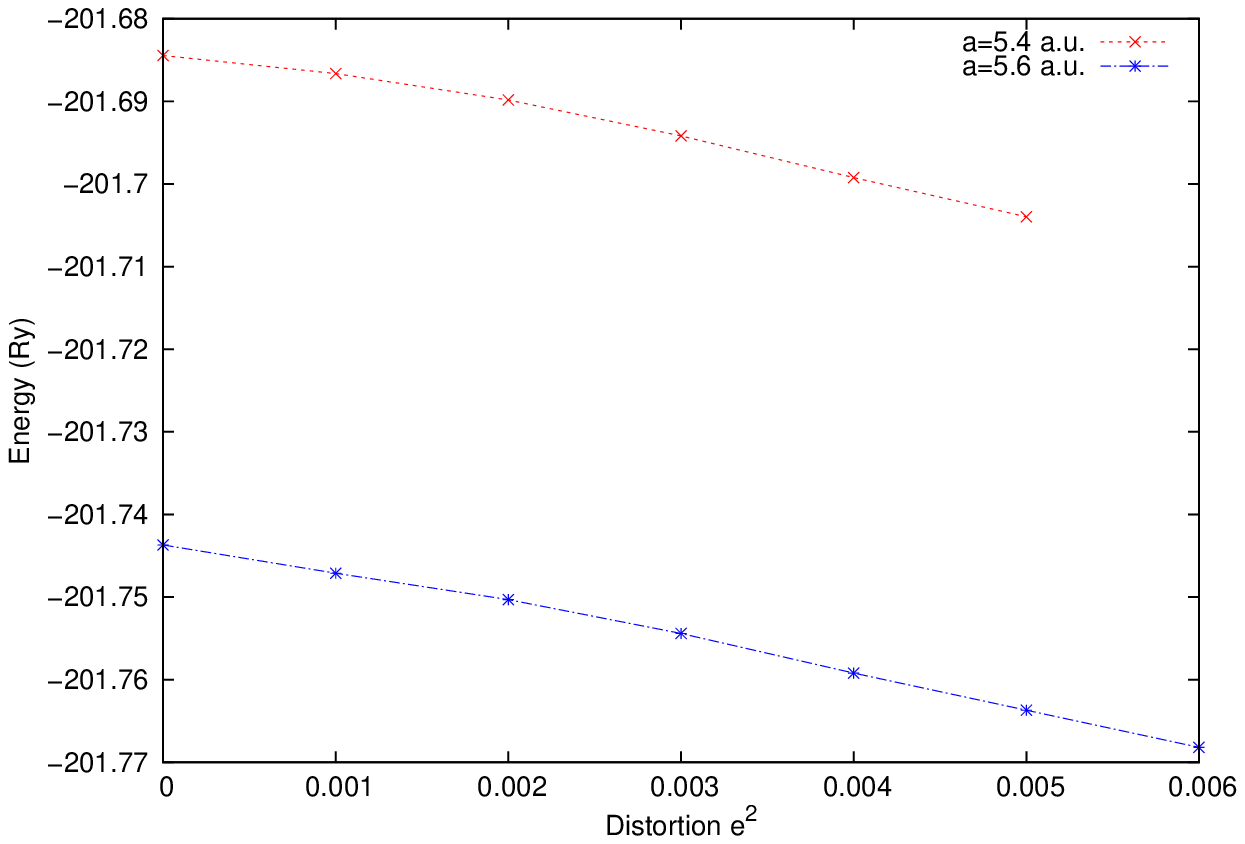}
}
\caption{(a) Energy v.\ Distortion Squared for a=5.4 and c44  (b) Energy v.\ Distortion Squared for a=5.6 and c44 and (c) Energy v.\ Distortiom Squared for c11-c12}
\end{figure}

It appears that the slope for c11-c12 has a small negative value suggesting 
an instability. So this result casts a doubt as to whether the H$_3$F can be 
a superconductor in the bcc-like structure. However, the unusually large values of the Hopfield parameter in the H-F system warrants further investigation in other crystal structures.

\section{Conclusion}
We emphasize that using the results of band structure calculations and
application of the GGM theory, the main conclusion of this work is that
H$_3$F has a very large value of the fluorine component of the Hopfield parameter. This is due to the very large
electron-ion matrix element $<I_f^2>$ on the fluorine site, and not to the $N(E_f)$, which has a
modest value similar to that in H$_3$S. However, due to an instability in
the calculated elastic constant c11-c12 in the Im$\bar{3}$m structure further studies are needed for other crystal structures to verify the present prediction.  

\section{Acknowledgments}
I acknowledge many useful discussions with Michael J. Mehl. This work was partially supported by DOE grant DE-SC0014337.


\newpage
\begin{thebibliography}{20}
\expandafter\ifx\csname natexlab\endcsname\relax\def\natexlab#1{#1}\fi
\expandafter\ifx\csname bibnamefont\endcsname\relax
  \def\bibnamefont#1{#1}\fi
\expandafter\ifx\csname bibfnamefont\endcsname\relax
  \def\bibfnamefont#1{#1}\fi
\expandafter\ifx\csname citenamefont\endcsname\relax
  \def\citenamefont#1{#1}\fi
\expandafter\ifx\csname url\endcsname\relax
 \def\url#1{\texttt{#1}}\fi
\expandafter\ifx\csname urlprefix\endcsname\relax\def\urlprefix{URL }\fi
\providecommand{\bibinfo}[2]{#2}
\providecommand{\eprint}[2][]{\url{#2}}

\bibitem[{\citenamefont{Ashcroft}(1968)}]{PhysRevLett.21.1748}
\bibinfo{author}{\bibfnamefont{N.~W.} \bibnamefont{Ashcroft}},
  \bibinfo{journal}{Phys. Rev. Lett.} \textbf{\bibinfo{volume}{21}},
  \bibinfo{pages}{1748} (\bibinfo{year}{1968}).

\bibitem[{\citenamefont{Papaconstantopoulos and
  Klein}(1977)}]{Ferro.16.307}
\bibinfo{author}{\bibfnamefont{D.~A.} \bibnamefont{Papaconstantopoulos}}
  \bibnamefont{and} \bibinfo{author}{\bibfnamefont{B.~M.} \bibnamefont{Klein}},
  \bibinfo{journal}{Ferroelectrics} \textbf{\bibinfo{volume}{16}},
  \bibinfo{pages}{307} (\bibinfo{year}{1977}),
  \eprint{http://dx.doi.org/10.1080/00150197708237185}.

\bibitem[{\citenamefont{Papaconstantopoulos
  et~al.}(1977)\citenamefont{Papaconstantopoulos, Boyer, Klein, Williams,
  Morruzzi, and Janak}}]{PhysRevB.15.4221}
\bibinfo{author}{\bibfnamefont{D.~A.} \bibnamefont{Papaconstantopoulos}},
  \bibinfo{author}{\bibfnamefont{L.~L.} \bibnamefont{Boyer}},
  \bibinfo{author}{\bibfnamefont{B.~M.} \bibnamefont{Klein}},
  \bibinfo{author}{\bibfnamefont{A.~R.} \bibnamefont{Williams}},
  \bibinfo{author}{\bibfnamefont{V.~L.} \bibnamefont{Morruzzi}},
  \bibnamefont{and} \bibinfo{author}{\bibfnamefont{J.~F.} \bibnamefont{Janak}},
  \bibinfo{journal}{Phys. Rev. B} \textbf{\bibinfo{volume}{15}},
  \bibinfo{pages}{4221} (\bibinfo{year}{1977}).

\bibitem[{\citenamefont{Gaspari and Gyorffy}(1972)}]{PhysRevLett.28.801}
\bibinfo{author}{\bibfnamefont{G.~D.} \bibnamefont{Gaspari}} \bibnamefont{and}
  \bibinfo{author}{\bibfnamefont{B.~L.} \bibnamefont{Gyorffy}},
  \bibinfo{journal}{Phys. Rev. Lett.} \textbf{\bibinfo{volume}{28}},
  \bibinfo{pages}{801} (\bibinfo{year}{1972}).

\bibitem[{\citenamefont{McMillan}(1968)}]{PhysRev.167.331}
\bibinfo{author}{\bibfnamefont{W.~L.} \bibnamefont{McMillan}},
  \bibinfo{journal}{Phys. Rev.} \textbf{\bibinfo{volume}{167}},
  \bibinfo{pages}{331} (\bibinfo{year}{1968}).

\bibitem[{\citenamefont{Drozdov et~al.}(2015)\citenamefont{Drozdov, Eremets,
  Troyan, Ksenofontov, and Shylin}}]{Drozdov2015}
\bibinfo{author}{\bibfnamefont{A.~P.} \bibnamefont{Drozdov}},
  \bibinfo{author}{\bibfnamefont{M.~I.} \bibnamefont{Eremets}},
  \bibinfo{author}{\bibfnamefont{I.~A.} \bibnamefont{Troyan}},
  \bibinfo{author}{\bibfnamefont{V.}~\bibnamefont{Ksenofontov}},
  \bibnamefont{and} \bibinfo{author}{\bibfnamefont{S.~I.}
  \bibnamefont{Shylin}}, \bibinfo{journal}{Nature}
  \textbf{\bibinfo{volume}{525}}, \bibinfo{pages}{73} (\bibinfo{year}{2015}),
  ISSN \bibinfo{issn}{0028-0836}, \bibinfo{note}{letter}.

\bibitem[{\citenamefont{Duan et~al.}(2014)\citenamefont{Duan, Liu, Tian, Li,
  Huang, Zhao, Yu, Liu, Tian, and Cui}}]{Duan2014}
\bibinfo{author}{\bibfnamefont{D.}~\bibnamefont{Duan}},
  \bibinfo{author}{\bibfnamefont{Y.}~\bibnamefont{Liu}},
  \bibinfo{author}{\bibfnamefont{F.}~\bibnamefont{Tian}},
  \bibinfo{author}{\bibfnamefont{D.}~\bibnamefont{Li}},
  \bibinfo{author}{\bibfnamefont{X.}~\bibnamefont{Huang}},
  \bibinfo{author}{\bibfnamefont{Z.}~\bibnamefont{Zhao}},
  \bibinfo{author}{\bibfnamefont{H.}~\bibnamefont{Yu}},
  \bibinfo{author}{\bibfnamefont{B.}~\bibnamefont{Liu}},
  \bibinfo{author}{\bibfnamefont{W.}~\bibnamefont{Tian}}, \bibnamefont{and}
  \bibinfo{author}{\bibfnamefont{T.}~\bibnamefont{Cui}},
  \bibinfo{journal}{Scientific Reports} \textbf{\bibinfo{volume}{4}},
  \bibinfo{pages}{6968} (\bibinfo{year}{2014}).

\bibitem[{\citenamefont{Papaconstantopoulos
  et~al.}(2015)\citenamefont{Papaconstantopoulos, Klein, Mehl, and
  Pickett}}]{PhysRevB.91.184511}
\bibinfo{author}{\bibfnamefont{D.}~\bibnamefont{Papaconstantopoulos}},
  \bibinfo{author}{\bibfnamefont{B.~M.} \bibnamefont{Klein}},
  \bibinfo{author}{\bibfnamefont{M.~J.} \bibnamefont{Mehl}}, \bibnamefont{and}
  \bibinfo{author}{\bibfnamefont{W.~E.} \bibnamefont{Pickett}},
  \bibinfo{journal}{Phys. Rev. B} \textbf{\bibinfo{volume}{91}},
  \bibinfo{pages}{184511} (\bibinfo{year}{2015}).

\bibitem[{\citenamefont{Bernstein et~al.}(2015)\citenamefont{Bernstein,
  Hellberg, Johannes, Mazin, and Mehl}}]{PhysRevB.91.060511}
\bibinfo{author}{\bibfnamefont{N.}~\bibnamefont{Bernstein}},
  \bibinfo{author}{\bibfnamefont{C.~S.} \bibnamefont{Hellberg}},
  \bibinfo{author}{\bibfnamefont{M.~D.} \bibnamefont{Johannes}},
  \bibinfo{author}{\bibfnamefont{I.~I.} \bibnamefont{Mazin}}, \bibnamefont{and}
  \bibinfo{author}{\bibfnamefont{M.~J.} \bibnamefont{Mehl}},
  \bibinfo{journal}{Phys. Rev. B} \textbf{\bibinfo{volume}{91}},
  \bibinfo{pages}{060511} (\bibinfo{year}{2015}).

\bibitem[{\citenamefont{Errea et~al.}(2015)\citenamefont{Errea, Calandra,
  Pickard, Nelson, Needs, Li, Liu, Zhang, Ma, and
  Mauri}}]{PhysRevLett.114.157004}
\bibinfo{author}{\bibfnamefont{I.}~\bibnamefont{Errea}},
  \bibinfo{author}{\bibfnamefont{M.}~\bibnamefont{Calandra}},
  \bibinfo{author}{\bibfnamefont{C.~J.} \bibnamefont{Pickard}},
  \bibinfo{author}{\bibfnamefont{J.}~\bibnamefont{Nelson}},
  \bibinfo{author}{\bibfnamefont{R.~J.} \bibnamefont{Needs}},
  \bibinfo{author}{\bibfnamefont{Y.}~\bibnamefont{Li}},
  \bibinfo{author}{\bibfnamefont{H.}~\bibnamefont{Liu}},
  \bibinfo{author}{\bibfnamefont{Y.}~\bibnamefont{Zhang}},
  \bibinfo{author}{\bibfnamefont{Y.}~\bibnamefont{Ma}}, \bibnamefont{and}
  \bibinfo{author}{\bibfnamefont{F.}~\bibnamefont{Mauri}},
  \bibinfo{journal}{Phys. Rev. Lett.} \textbf{\bibinfo{volume}{114}},
  \bibinfo{pages}{157004} (\bibinfo{year}{2015}).

\bibitem[{flo()}]{flores}
\bibinfo{note}{J. A. Flores-Livas, A. Sanna, and E. K. U. Gross, The European
  Physical Journal B89 (3), 1-6.}

\bibitem[{\citenamefont{Quan and Pickett}(2016)}]{quan}
\bibinfo{author}{\bibfnamefont{Y.}~\bibnamefont{Quan}} \bibnamefont{and}
  \bibinfo{author}{\bibfnamefont{W.~E.} \bibnamefont{Pickett}},
  \bibinfo{journal}{Phys. Rev. B} \textbf{\bibinfo{volume}{93}},
  \bibinfo{pages}{104526} (\bibinfo{year}{2016}).

\bibitem[{\citenamefont{Li et~al.}(2014)\citenamefont{Li, Hao, Liu, Li, and
  Ma}}]{li}
\bibinfo{author}{\bibfnamefont{Y.}~\bibnamefont{Li}},
  \bibinfo{author}{\bibfnamefont{J.}~\bibnamefont{Hao}},
  \bibinfo{author}{\bibfnamefont{H.}~\bibnamefont{Liu}},
  \bibinfo{author}{\bibfnamefont{Y.}~\bibnamefont{Li}}, \bibnamefont{and}
  \bibinfo{author}{\bibfnamefont{Y.}~\bibnamefont{Ma}}, \bibinfo{journal}{J.
  Chem. Phys} \textbf{\bibinfo{volume}{140}}, \bibinfo{eid}{174712}
  (\bibinfo{year}{2014}).

\bibitem[{bia()}]{bianconi}
\bibinfo{note}{A. Bianconi and T. Jarlborg, Novel Superconducting Materials, Vol 1, 
  Issue 1, Issue1, ISSN (Online) 2299-3193.}

\bibitem[{fan()}]{fan}
\bibinfo{note}{F. Fan, D.A. Papaconstantopoulos, M.J. Mehl, and B.M. Klein, Journal of
Physics and Chemistry of Solids, 99, 105-110 (2016).}
 
\bibitem[{\citenamefont{Singh}(1994)}]{singh}
\bibinfo{author}{\bibfnamefont{D.~J.} \bibnamefont{Singh}},
  \emph{\bibinfo{title}{Plane waves, pseudopotentials, and the LAPW Method}}
  (\bibinfo{publisher}{Kluwer Academic Publishers}, \bibinfo{address}{Boston},
  \bibinfo{year}{1994}).

\bibitem[{nrl()}]{nrl}
\bibinfo{note}{The NRL LAPW code, originally developed by H. Krakauer and D. J.
  Singh, was used with Hedin-Lundqvist exchange-correlation. DOS results were
  generated from 1785 k points in the irreducible Brillouin zone with the
  tetrahedron method. Total energies were fit to the Birch equation to obtain
  the P(V) equation of state.}

\bibitem[{\citenamefont{Birch}(1978)}]{birch}
\bibinfo{author}{\bibfnamefont{F.}~\bibnamefont{Birch}},
  \bibinfo{journal}{Journal of Geophysical Research: Solid Earth}
  \textbf{\bibinfo{volume}{83}}, \bibinfo{pages}{1257} (\bibinfo{year}{1978}),
  ISSN \bibinfo{issn}{2156-2202}.

\bibitem[{\citenamefont{Allen and Dynes}(1975)}]{PhysRevB.12.905}
\bibinfo{author}{\bibfnamefont{P.~B.} \bibnamefont{Allen}} \bibnamefont{and}
  \bibinfo{author}{\bibfnamefont{R.~C.} \bibnamefont{Dynes}},
  \bibinfo{journal}{Phys. Rev. B} \textbf{\bibinfo{volume}{12}},
  \bibinfo{pages}{905} (\bibinfo{year}{1975}).

\bibitem[{\citenamefont{Bennemann and Garland}(1972)}]{bennemann}
\bibinfo{author}{\bibfnamefont{K.~H.} \bibnamefont{Bennemann}}
  \bibnamefont{and} \bibinfo{author}{\bibfnamefont{J.~W.}
  \bibnamefont{Garland}},\bibinfo{journal}{AIP Conf. Proc.}
  \textbf{\bibinfo{volume}{4}},\bibinfo{pages}{103} (\bibinfo{year}{1972}).


\end{thebibliography}
\end{document}